\documentclass[epsfig,graphics,twocolumn,floatfix,mathbbm,pra]{revtex4}
\usepackage{graphicx}
\usepackage[hidelinks]{hyperref}
\usepackage{amsmath,bbm}
\usepackage{amssymb}

\begin{document}

\title[a]{Noise in optical quantum memories based on dynamical decoupling of spin states}%

\author{Emmanuel Zambrini Cruzeiro}
\author{Florian Fr\"{o}wis}
\email{florian.froewis@unige.ch}
\author{Nuala Timoney}
\author{Mikael Afzelius}
\address{Group of Applied Physics, University of Geneva, CH-1211 Geneva, Switzerland}%

\date{\today}

\begin{abstract}
Long-lived optical quantum memories are of great importance for scalable distribution of entanglement over remote networks (e.g.~quantum repeaters). Long-lived storage generally relies on storing the optical states as spin excitations since these often exhibit long coherence times. To extend the storage time beyond the intrinsic spin dephasing time one can use dynamical decoupling techniques. However, it has been shown that dynamical decoupling introduces noise in optical quantum memories based on ensembles of atoms. In this article a simple model is proposed to calculate the resulting signal-to-noise ratio, based on intrinsic quantum memory parameters such as the optical depth of the ensemble. We also characterize several dynamical decoupling sequences that are efficient in reducing this particular noise. Our calculations indicate that it should be feasible to reach storage times well beyond one second under reasonable experimental conditions.

\end{abstract}

\maketitle

\section{INTRODUCTION}

Optical quantum memories are stationary devices that are able to store quantum states of light for retrieval at a later point in time \cite{Lvovsky2009}. These can be used to synchronize probabilistic quantum optics processes, which is crucial for the scalability of many optical quantum technologies \cite{Bussieres2013}. A prominent example is the DLCZ-type quantum repeater \cite{Duan2001,Sangouard2011} for distributing entanglement over large distances, in which quantum memories are used to store excitations entangled with propagating photons. For quantum repeaters, or large-scale optical quantum networks in general \cite{Kimble2008}, one requires long-lived quantum memories \cite{Collins2007,Razavi2009}. In a DLCZ-type quantum repeater the memory lifetime must be longer than the average time to distribute entanglement over the entire repeater length $L$ \cite{Sangouard2011}, which emphasizes the importance of long-lived quantum memories. 

Atomic ensembles can be used to create efficient quantum memories thanks to the strong collective enhancement of the light-matter interaction \cite{Hammerer2010}. In addition one can use spin states for long-duration storage as these have long coherence times \cite{Pascual-Winter2012,Heinze2014,Arcangeli2014,Dudin2013,Zhong2015,Radnaev2010}. In ensembles the spin states are generally subject to dephasing processes, due to inhomogeneous spin broadening and/or due to coupling to the environment. These processes are characterized by the $T_2^{*}$ and $T_2$ times, respectively. The dephasing causes a strong loss in the optical read-out efficiency of the memory. To counter this dephasing one can apply spin echo techniques. A simple Hahn echo sequence, which employs a single population-inversion pulse (e.g. a $\pi$ pulse), can completely undo the inhomogeneous dephasing. The memory time is then limited by the dephasing due to the environment, i.e., by $T_2$. However, even the dephasing time $T_2$ can be increased using dynamical decoupling (DD) sequences \cite{Viola1998}, which employ series of population-inversion pulses \cite{Lange2010,Bylander2011,Souza2011,Wang2012b,Heinze2014}. The resulting, effective dephasing time $T_2^{\mathrm{DD}}$ depends on a range of factors, such as the spacing between the pulses $\tau$ with respect to the correlation time $\tau_{\mathrm{c}}$ of the dephasing process \cite{Pascual-Winter2012,Arcangeli2014}, errors in the pulses \cite{Heshami2011,Souza2011} and the spectrum of the dephasing noise \cite{Biercuk2009,Bylander2011,Alvarez2011}.

\begin{figure*}
    \centering
    \includegraphics[width=.95\textwidth]{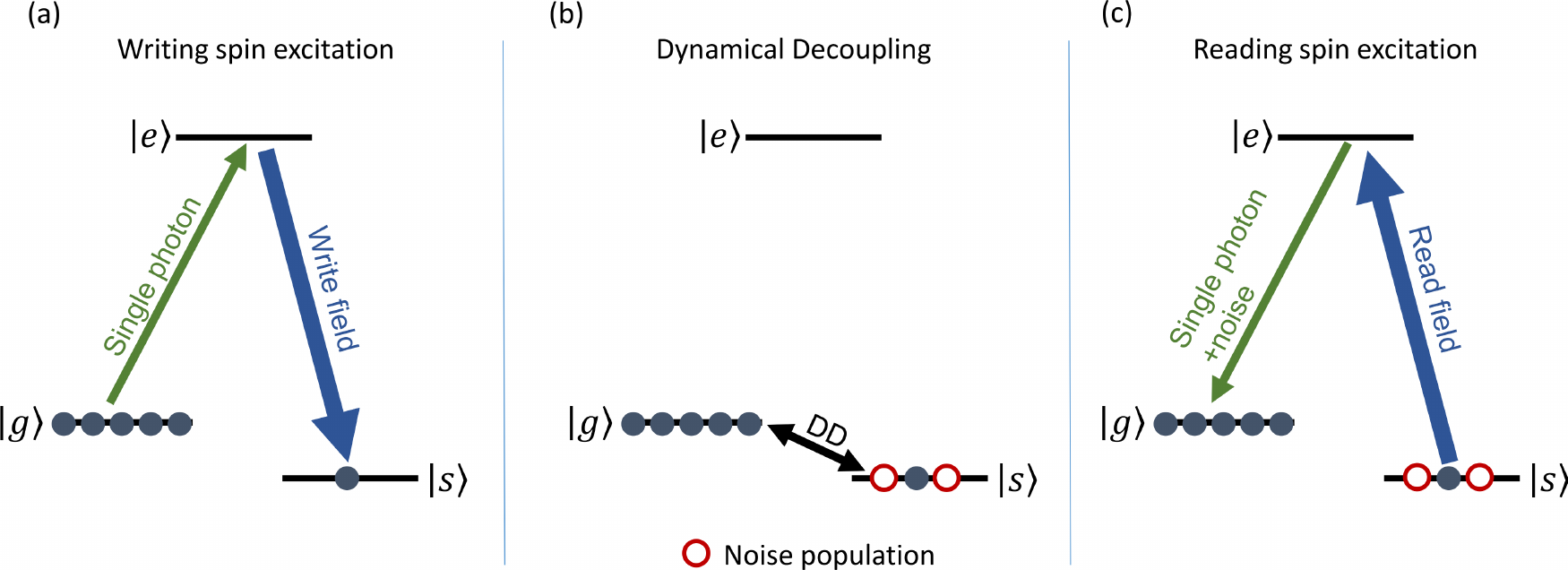}
    \caption{Basic scheme of a universal quantum memory based on storage of an optical single photon as a single spin excitation. The scheme includes dynamical decoupling (DD) of the spin states to increase the storage time. In the energy diagram $|g\rangle$ and $|s\rangle$ are two spin states in the electronic ground state, while $|e\rangle$ is an electronic excited state in the optical regime. Before storage all atoms are spin polarized into $|g\rangle$. a) The first step is to write a single spin excitation into $|s\rangle$, which here is generated through storage of a single photon using a strong classical write field. Alternatively one can use the DLCZ scheme to generate a single spin excitation \cite{Duan2001}. b) The single spin excitation is manipulated by a DD sequence comprised of a large, even number of population inversion $\pi$ pulses. Pulse area errors in the $\pi$ pulses will cause a relative increase of the population in $|s\rangle$ shown as open circles. c) The spin excitation is converted into a single photon using a strong read field, which will also cause spontaneous emission noise due to the additional population in $|s\rangle$ caused by the imperfect DD sequence.}
   \label{FIG_optical_storage_sequence}
\end{figure*}

Errors in the population-inversion pulses generally reduce the coherence of the stored spin excitation, which in turn causes a reduction of the effective optical storage time. Even more importantly the errors cause optical noise when reading out the memory \cite{Heshami2011}, which reduces the fidelity of the storage process. To understand the origin of this particular kind of noise we consider a basic atomic three-level system in a $\Lambda$ configuration where two spin states $|g\rangle$ and $|s\rangle$ are optically coupled to a common excited state $|e\rangle$, see Fig.~\ref{FIG_optical_storage_sequence}. We assume that a single spin excitation has been generated in $|s\rangle$, delocalized over all atoms, which is described by a non-symmetric Dicke state $|W_s\rangle$=$\sum_k c_k |g\cdot\cdot\cdot s_k \cdot\cdot\cdot g\rangle$ where $c_k$ are amplitudes.  The generation of the spin excitation can be done using some optical storage scheme which converts a single optical photon into a single spin excitation \cite{Lvovsky2009}, or by spontaneous Raman scattering where a detection of a Stokes photon heralds a single spin excitation as in the DLCZ scheme \cite{Duan2001}.

The population-inversion pulses ($\pi$ pulses) of the spin echo sequence then swap the population between the $|g\rangle$ and $|s\rangle$ states at each time interval $\tau$. To restore the initial $|W_s\rangle$ state an even number of pulses is used. The optical read-out is done by applying a strong control field on $|s\rangle$-$|e\rangle$, which converts the single spin excitation to a single photon in the $|g\rangle$-$|e\rangle$ mode \cite{Duan2001,Gorshkov2007c,Afzelius2009a,Sekatski2011}. Imperfect population-inversion pulses will, however, cause additional excitations in $|s\rangle$, which will create spontaneous emission noise during the optical read-out \cite{Johnsson2004,Heshami2011}, as illustrated in  Fig.~\ref{FIG_optical_storage_sequence}. Heshami \textit{et al.}~\cite{Heshami2011} studied this intrinsic noise source for a simple Hahn echo sequence. They showed that the noise can be sufficiently suppressed for small area errors of the $\pi$ pulse (typically around 1\%). Recently two optical storage experiments have confirmed that spin-echo manipulation of a single spin excitation in an ensemble is indeed possible without introducing excessive noise \cite{Jobez2015,Rui2015}. Heshami \textit{et al.} also made a short calculation of the required pulse area precision for a DD sequence, however only for the most basic one. We emphasize that DD sequences have not yet been tested experimentally in this context.

In this article we analyze in detail several DD sequences for extending the memory time of a single-excitation quantum memory based on an ensemble of atoms. For this, we first develop a realistic model for calculating the resulting signal-to-noise ratio (SNR) of the memory read-out in Sec.~\ref{sec:QM_performance}. The model is sufficiently general to treat several quantum memory schemes such as electromagnetically induced transparency (EIT) \cite{Phillips2001}, gradient echo memory (GEM) \cite{Hetet2008a,Hedges2010} and atomic frequency comb (AFC) \cite{Afzelius2009a} memories. We then study the SNR for several well-known DD sequences, which are introduced in Sec.~\ref{sec:DD_sequences}. Neglecting first homogeneous broadening in Sec.~\ref{sec:coher-popul-dd}, we analytically calculate the relative amount of extra population induced in the spin state $|s\rangle$ (noise) and the loss in the coherence of the $|W_s\rangle$ state (memory efficiency) due to imperfect population inversion by the spin echo pulses. We find that, for all DD sequences we considered, the SNR is limited by the photon noise caused by extra population, while the loss in memory efficiency is negligible in the regime where SNR $\gg$ 1. The results are compared to numerical simulations. These findings confirm and extend the results of Ref.~\cite{Heshami2011}. Using more sophisticated sequences with higher robustness generally reduces the noise and increases the effective storage time $T_2^{\mathrm{DD}}$ for which a high SNR can be obtained. However, our results also show that there is no magical sequence that will limit this noise particularly well.
In section \ref{sec:Numerical}, we explicitly take into account the influence of homogeneous broadening and present arguments for the optimal delay time between the $\pi$ pulses. In addition, we shortly discuss the influence of imperfect phase relations between the pulses.
Finally, we discuss the potential of reaching long storage times with realistic experimental parameters in Sec. \ref{sec:Conclusions_Outlook}.

\section{SPIN-ENSEMBLE BASED QUANTUM MEMORY PERFORMANCE}
\label{sec:QM_performance}

A universal read-write quantum memory, viewed as a black box, is a memory in which a quantum state of light can be stored and later retrieved on demand. The quantum state is often carried by a single photon, which is stored as a spin excitation in an atomic memory, see Fig.~\ref{FIG_optical_storage_sequence}. For a read-write memory the memory efficiency $\eta_M$ is the overall efficiency to both write and read the memory, including a potential intra-memory loss due to dephasing. For all quantum memories based on an ensemble of atoms, the efficiency is a function of the optical depth $\tilde{d}$ of the memory material \cite{Hammerer2010}. As an example, for memory schemes that are based on dephasing and rephasing of an inhomogeneously broadened ensemble, such as the gradient echo memory (GEM) \cite{Sangouard2007,Sparkes2013} or the atomic frequency comb (AFC) memory \cite{Afzelius2009a}, the memory efficiency can be calculated as

\begin{equation}\label{eq:2.2}
\eta_M = (1-e^{-\tilde{d}})^2\eta_D,
\end{equation}

\noindent where $\widetilde{d}$ is an effective optical depth and $\eta_D$ accounts for loss in efficiency due to dephasing (independent of the optical depth and not related to DD). Note that for AFC memories this formula holds for backward read-out, while for forward read-out the efficiency is $\eta_M = \tilde{d}^2 e^{-\tilde{d}} \eta_D$. For the popular memory scheme based on electromagnetically induced transparency (EIT), the efficiency has also been calculated as a function of optical depth \cite{Gorshkov2007c}. Generally it is $\eta_M = 1-C/\tilde{d}$ for optimal storage in the limit of large optical depth $\tilde{d}$, where $C$ is a constant. 

We now consider the effects of the DD sequence on the storage process. The initial spin state before applying the DD sequence is taken to be 

\begin{equation}\label{eq:2}
|W_s\rangle=\sum_{k=1}^{N} c_k |g\cdot\cdot\cdot s_k \cdot\cdot\cdot g\rangle,
\end{equation}

\noindent where $N$ is the number of atoms and $c_k$ is a probability amplitude. The DD sequence can be represented by a unitary transformation $\mathcal{T}^{\mathrm{DD}}=T^{\mathrm{DD}}_1 \otimes \dots \otimes T^{\mathrm{DD}}_N$, where the index indicates that the unitaries are different for every spin, such that the final state after the DD sequence is $\mathcal{T}^{\mathrm{DD}} |W_s\rangle$. Errors in the DD sequence transform the initial state such that the final state cannot be read-out optically with the same efficiency. We denote the efficiency of the DD sequence as $\eta_{\mathrm{coh}}$, such that the overall memory efficiency including the DD sequence is $\eta_M \eta_{\mathrm{coh}}$. 

Imperfections can also induce extra population in the $|s\rangle$ state, which we denote by the fractional population term $\rho_{\mathit{ss}}$. If we assume that this population is completely transferred to the excited state by the optical read-out field it will cause spontaneous emission in the output mode. Following the calculations in Ref.~\cite{Ledingham2010,Sekatski2011} it can be shown that the average number of photons spontaneously emitted into the output mode is simply given by 

\begin{equation}
\mu_{noise} = (1-e^{-\widetilde{d}})\rho_{\mathit{ss}},
\end{equation}

\noindent provided that $\mu_{noise} \ll 1$.

If we now consider storage of an input mode, with $\mu_{in}$ number of photons in average in the mode, the SNR in the output mode is then calculated as

\begin{equation} \label{eq:4}
	\mathrm{SNR} = \frac{\mu_{in}\eta_M \eta_{\mathrm{coh}}}{\mu_{noise}}=\frac{\mu_{in}\eta_M}{(1-e^{-\widetilde{d}})} \frac{\eta_{\mathrm{coh}}}{\rho_{\mathit{ss}}}.
\end{equation}

The performance of different DD sequences is thus reflected in the ratio  $R=\eta_{\mathrm{coh}}/\rho_{\mathit{ss}}$. The question then arises how to calculate $\eta_{\mathrm{coh}}$ and $\rho_{\mathit{ss}}$ for a given unitary $\mathcal{T}^{\mathrm{DD}}$ describing a particular DD sequence.

If we assume that, instead of storing a single photon Fock state $|1\rangle$, we store a coherent state $|\alpha\rangle$, with $\alpha = e^{i \beta}$, then the resulting state is the spin-coherent state $| \psi_{\mathrm{in}} \rangle ^{\otimes N}$. If we also assume, for simplicity, that the transfer probability to the spin state is without loss, then $|\psi_{\mathrm{in}}\rangle$ can be written as

\begin{equation}\label{eq:7}
|\psi_{\mathrm{in}}\rangle=\sqrt{1-\frac{1}{N}}|g\rangle+e^{i \beta}\frac{1}{\sqrt{N}}|s\rangle.
\end{equation}

\noindent In this case $\eta_{\mathrm{coh}}$ is proportional to the averaged coherence 
$|\rho_{\mathit{gs}}|^2 = \langle \sigma_x \rangle_{\bar{\rho}}^2 + \langle \sigma_y \rangle_{\bar{\rho}}^2$
where $\bar{\rho} = N^{-1}\sum_k T^{\mathrm{DD}}_k\left| \psi_{\mathrm{in}} \right\rangle \left\langle \psi_{\mathrm{in}} \right| T^{\mathrm{DD}\dagger}_k$ is the averaged state after the evolution $\mathcal{T}^{\mathrm{DD}}$ and $\sigma_x, \sigma_y, \sigma_z$ are the Pauli matrices. This follows from a semiclassical approach, where the signal strength is proportional to the expectation value of the dipole moment operator.

Storing a single photon $|1\rangle$ results in an entangled state $|W_s\rangle$. This state gives rise to difficulties in terms of the proper choice for $\eta_{\mathrm{coh}}$ \cite{Scully2006,Svidzinsky_Quantum_2015} and complexity of the calculations. As shown in more details in the appendix, we can circumvent these problems using spin coherent states, Eq.~(\ref{eq:7}). To summarize, we first note that the phase information of an arbitrary (entangled) state may not be given by the off-diagonal elements of the reduced one-particle state like for spin-coherent states. In contrast, the two-body correlation function is the relevant parameter for the inter-atomic phase coherence for arbitrary states. Hence, the reduced two-body states of $\mathcal{T}^{\mathrm{DD}} \left| W_s \right\rangle $ (or of any general state) determine $\eta_{\mathrm{coh}}$. However, it is straightforward to show that the reduced two-body states of $\mathcal{T}^{\mathrm{DD}} \left| W_s \right\rangle $ and $\mathcal{T}^{\mathrm{DD}} \left| \psi_{\mathrm{in}} \right\rangle ^{\otimes N}$ --averaged over $\beta$-- differ only by a $O(1/N^2)$ correction. With additional arguments, this remains even true if the amplitudes $c_k$ in Eq.~(\ref{eq:2}) differ from $N^{-1/2}$ (see appendix). Hence, we can approximate the two-body correlations very well by using spin-coherent states (\ref{eq:7}) and integrate over $\beta$. We then find that 
\begin{equation}
\label{eq:3}
\eta_{\mathrm{coh}} = \int_0^{2\pi} N \frac{d\beta}{2\pi} \left( \langle \sigma_x \rangle_{\bar{\rho}}^2 + \langle \sigma_y \rangle_{\bar{\rho}}^2 \right).
\end{equation}
Similar considerations allow us to use $| \psi_{\mathrm{in}} \rangle ^{\otimes N}$ instead of $| W_s \rangle $ to quantify the generated noise. One finds that
\begin{equation}
\label{eq:5}
\rho_{\mathit{\mathit{ss}}} = \int_0^{2\pi} \frac{d\beta}{2\pi} \left\langle s \right| \bar{\rho} \left| s \right\rangle -\frac{1}{N},
\end{equation}
where the $1/N$ corrects the contribution from the input signal.

\section{Dynamical decoupling sequences}
\label{sec:DD_sequences}

In this section, we first introduce the general pattern of a DD sequence and some well-known examples. Next, we discuss some basic features of these sequences. Finally, we comment on composite pulses \cite{tycko1983broadband,tycko1985fixed} and other more elaborate sequences \cite{Gullion1990}.

\subsection{General scheme of a DD sequence}
\label{sec:general-scheme-dd}

\begin{figure}[htbp]
\centerline{\includegraphics[width=\columnwidth]{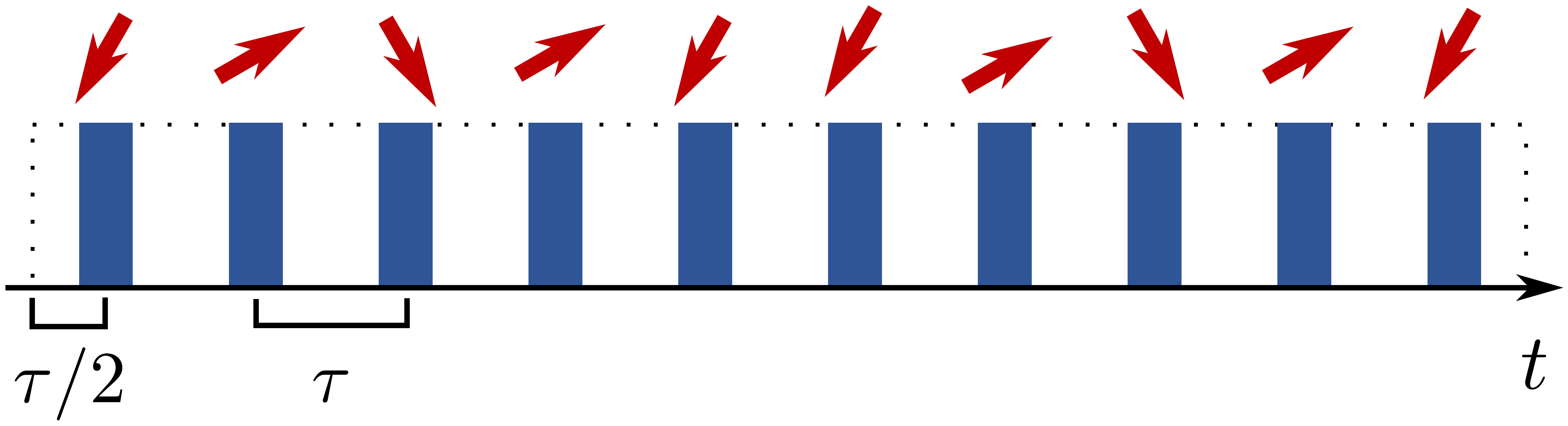}}
\caption[]{\label{fig:schematics} Schematic of a single repetition of a DD sequence (dotted box). At times separated by $\tau$, instantaneous $\pi$ pulses are applied with different phases $\varphi$ (symbolized by the red arrows). One such block is then repeated $m$ times. Here, the example of U5a:CP is shown [see Sec.~\ref{sec:comb-comp-puls} and Eq.~(\ref{eq:15})], which is a ten-pulse sequence. Note that we introduce the same time gaps between pulses \textit{within} composite pulses.}
\end{figure}

DD sequences treated in this paper follow a simple pattern (see Fig.~\ref{fig:schematics} for a schematic). Free evolution is described by
\begin{equation}\label{eq:8}
  V_{\tau}=\cos \left(\Delta \tau/2\right)\mathbbm{1}-i \sin \left(\Delta \tau/2\right)\sigma_z 
\end{equation}
where $\Delta$ is the detuning of a given spin, that is, the distance from the center of absorption in the frequency domain. For the moment, we assume that the detuning is constant in time, while the detunings of the spins follow some inhomogeneous distribution. A time-dependent (stochastic) fluctuation of $\Delta$ (i.e., homogeneous broadening) is discussed later in Sec.~\ref{sec:Numerical}.
In addition, spin ensembles can be manipulated using $\pi$ pulses, just like in spin-echo techniques. We treat light-matter interaction with a semi-classical model based on the Jaynes-Cummings Hamiltonian \cite{jaynes1963comparison} for interaction of light with a two-level system, applied to ensembles of spins. It is important to take pulse errors into account. In this paper, we restrict ourselves to systematic errors  $\epsilon$ in amplitude. 
The propagator that describes the application of an instantaneous pulse with phase $\varphi$ and amplitude $\pi+\epsilon$ on a single ion is given by
\begin{equation}\label{eq:9}
\Pi_{\varphi}= \cos(\epsilon/2) (\cos \varphi\, \sigma_x + \sin \varphi\, \sigma_y) + i \sin(\epsilon/2) \mathbbm{1}.
\end{equation} Experimentally, small values of $\epsilon$ can be achieved. For example, the amplitude error per pulse was estimated to be around $\epsilon \approx 0.06 \pi$ in Ref. \cite{Jobez2015}.

With Eqs.~(\ref{eq:8}) and (\ref{eq:9}), one can build up arbitrary $\pi$-pulse-based sequences. The basic mechanism of DD is easily illustrated for perfect population inversion since 
\begin{equation}\label{eq:10}
V_{\tau}\Pi_{\varphi}V_{\tau}\bigg|_{\epsilon =0}=\Pi_{\varphi}
\end{equation}
holds for any $\tau$ when there are no errors, which means that the effect of inhomogeneous broadening is corrected.
Note that Eq.~(\ref{eq:10}) suggests to use only a single application of a $\pi$ pulse (as done in the Hahn echo) to counter inhomogeneous broadening. However, we implicitly take homogeneous broadening into account, which entails that $\tau$ has to be much smaller than the correlation time $\tau_{\mathrm{c}}$ of the dephasing process \cite{Pascual-Winter2012,Arcangeli2014} (see Sec.~\ref{sec:Numerical} for more details). Hence, one has to apply many $\pi$ pulses to reach long coherence times. To summarize, we call a DD sequence a series of $\pi$ pulses $\Pi(\varphi)$ with potentially different phases $\varphi$. Between the pulses, free evolution takes place (see Fig.~\ref{fig:schematics}). To simplify defining future sequences, we abbreviate the basic building block of any sequence with the unitary
\begin{equation}
\label{eq:11}
U(\varphi) = V_{\tau/2}\Pi_{\varphi}V_{\tau/2}.
\end{equation}
A single repetition of a DD sequence reads $T^{\mathrm{DD}} = U(\varphi_L)\dots U(\varphi_1)$, where $L$ is the length of a sequence. Without errors, one always has $T^{\mathrm{DD}}|_{\epsilon=0}= \mathbbm{1}$ (up to global phases). To reach longer storage times $t$, we repeat the sequence $m$ times such that $t = \tau L m$.

\subsection{Examples of simple DD sequences}
\label{sec:examples-simple-dd}

The most basic dynamical decoupling sequence is the Carl-Purcell (CP) sequence \cite{carr1954effects}. It consists of two $\pi$ pulses with zero phase
\begin{equation}\label{eq:CP}
T^{\mathrm{CP}}= U(0)^{2}.
\end{equation}
The Carl-Purcell-Meiboom-Gilles (CPMG) sequence was introduced to compensate first-order amplitude errors for certain input states \cite{meiboom1958modified}. Compared to Eq.~(\ref{eq:CP}), the second pulse is of opposite phase
\begin{equation}
T^{\mathrm{CPMG}}= U(\pi)U(0).
\end{equation}
As we will see later, these two sequences behave equivalently for the phase-averaged spin-coherent states we use. Hence, we will only consider the CP sequence in the following. The simplest sequence which partially compensates amplitude errors for any initial state is the XY4 sequence \cite{maudsley1986modified,Souza2011,Wang2012b}. It is defined via an alternation of pulses in $x$ and $y$ directions and reads
\begin{equation}\label{eq:13}
  T^{\mathrm{XY4}}= U(\pi/2)U(0)U(\pi/2)U(0)
\end{equation}

\label{sec:dd-sequ-pres}

Before we introduce additional sequences, we study the basic properties of CP, CPMG and XY4. Since any two-level unitary operation is isomorphic to a rotation in a three-dimensional real space, it can be written in the form
\begin{equation}
\label{eq:12}
T^{\mathrm{DD}} = \exp\left(- \frac{i}{2} \alpha \vec{n} \cdot \vec{\sigma}\right)
\end{equation}
up to a global phase, where $\vec{n}$ is the unit vector that points in direction of the rotation axis and $\alpha$ is the angle that determines how much $T^{\mathrm{DD}}$ rotates. In our case, $\vec{n}$ and $\alpha$ are functions of $\tau$, $\Delta$, $\epsilon$ and the phases $\varphi$ of the specific sequences. The advantage of this representation is the intuitive account to understand the action of the sequence. Furthermore, many repetitions $m$ of the same sequence just change the angle of $T^{\mathrm{DD}}$ from $\alpha$ to $m \alpha$.

\begin{figure}[htbp]
\centerline{\includegraphics[width=0.7\columnwidth]{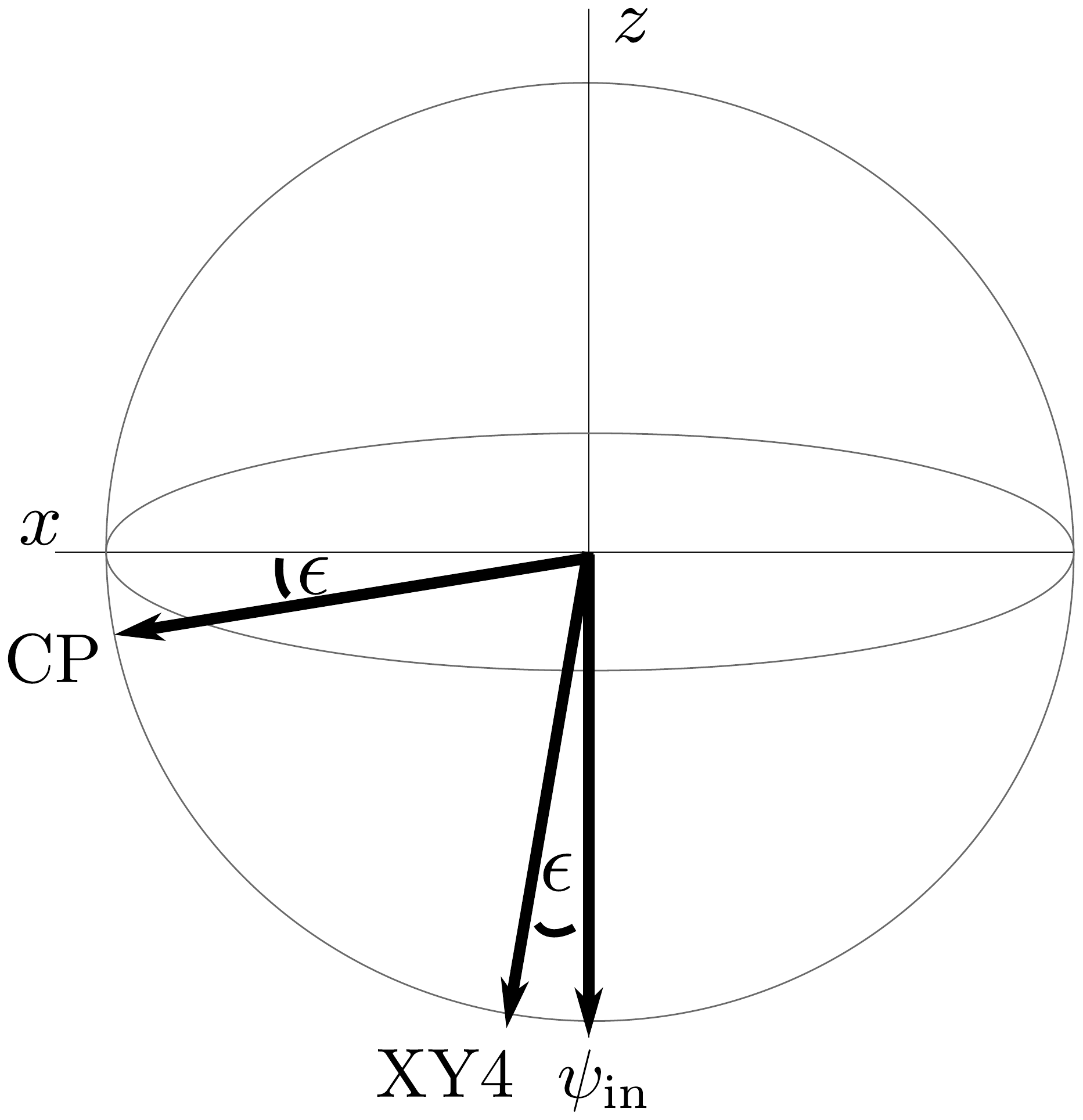}}
\caption[]{\label{fig:Bloch}  Sketch of Bloch sphere representations of the initial state $| \psi_{\mathrm{in}} \rangle $, Eq.~(\ref{eq:7}), and the sequences CP, Eq.~(\ref{eq:CP}), and XY4, Eq.~(\ref{eq:13}). Since $N\gg 1$, the initial state is very close to the ground state. The rotation axis of $T_{\mathrm{CP}}$ is $\epsilon$-close to the $x$ axis, while $T_{\mathrm{XY4}}$ is $\epsilon$-close to the $z$ axis.}
\end{figure}

It turns out that, for CP, CPMG and XY4, $\vec{n}$ and $\alpha$ reduce to simple functions in the limit of small amplitude errors $\epsilon \ll 1$. For even simpler expressions, we represent $\vec{n}$ in spherical coordinates $(\theta,\phi)$. The results are summarized in Table \ref{table:parameters2} and sketched in Fig.~\ref{fig:Bloch}. We observe that $\vec{n}$ lies close to the equator for CP and CPMG and that only their azimuthal angles differ. This shows that CP and CPMG are equivalent for our problem since we phase-average over the population and coherence in Eqs.~(\ref{eq:5}) and (\ref{eq:3}), respectively. In contrast, the XY4 sequence rotates around an axis that is close to the $z$ axis. This implies that the population generated by XY4 is bounded, where the bound is roughly given by the square of polar angle $\theta^2 \propto \epsilon^{2}$. In contrast, if $T^{\mathrm{DD}}$ rotates around an axis close to the equator (like CP), population is generated without a nontrivial bound. In addition, the angle $\alpha$ for XY4 is quadratically suppressed compared to CP and CPMG. This is a manifestation of the vanishing first-order contribution of the amplitude errors in XY4 \cite{Wang2012b}.

  \begin{table}[ht]\setlength{\tabcolsep}{0.2 cm}
    \centering
    \begin{tabular}{l l l l l} 
      \hline\hline 
      Sequence &$\alpha$& $\theta$ & $\phi$ & $L$ \\ 
      \hline 
      CP & $ 2c_1\epsilon$ & $\frac{\pi}{2}- \frac{1}{2}s_1\epsilon$ & 0 & 2\\ 
      CPMG & $ 2c_1\epsilon$ & $\frac{\pi}{2}- \frac{1}{2}s_1\epsilon$ & $\frac{\pi}{2}$ & 2\\ 
      XY4 &$ c_2\epsilon^2$& $ \frac{1}{\sqrt{2}} (c_1+s_1)\epsilon$ & $\frac{\pi}{4}$& 4 \\ 
      XY8 &$\frac{1}{\sqrt{2}} (c_1 +c_3)\epsilon^3$& $ \frac{\pi}{2} - s_1 \epsilon$ & $\frac{5\pi}{4} + \frac{1}{2}c_2\epsilon^2 $ & 8\\
      U5a:CP &$O(\epsilon^3)$& $\frac{\pi}{2} + O(\epsilon^3) $ & $ O(\epsilon^2)$ &10 \\
      U5a:XY4 &$O(\epsilon^6)$& $ O(\epsilon^3) $ & $\frac{\pi}{4} + O(\epsilon^5)$&20 \\
      \hline \hline
    \end{tabular}
    \caption{\label{table:parameters2}
Approximations for the parameters of Eq.~(\ref{eq:12}) for several sequences introduced in the text and the number $L$ of building blocks of type Eq.~(\ref{eq:11}). The parameters are given in spherical coordinates $n_x = \sin \theta \cos \phi$,  $n_y = \sin \theta \sin \phi$ and $n_z = \cos \theta$ using the abbreviations $c_k \equiv \cos(\frac{1}{2}k \Delta \tau)$ and $s_k \equiv \sin(\frac{1}{2}k \Delta \tau)$. The approximations are valid in the limit $\epsilon \ll 1$. U5a:CP and U5a:XY4 denote sequences where in a CP and XY4 sequence, respectively, a $\pi$ pulse $U(\varphi)$ is replaced by a composite pulse [see Eq.~(\ref{eq:15})]. The parameters of U5a:CP and U5a:XY4 are more complicated and thus only the scaling of leading order in $\epsilon$ is mentioned here.}
  \end{table}

\subsection{More sophisticated DD sequences and composite pulses}\label{sec:comb-comp-puls}

In this section, we comment on more elaborate pulse sequences like XY8 \cite{Gullion1990} and KDD \cite{Souza2011}, including composite pulses \cite{Levitt_Composite_1986,Genov_Correction_2014} and finally establish links between them. 
We previously compared XY4 and CP. In XY4, $\alpha$ is quadratically suppressed (see Table \ref{table:parameters2}), implying that XY4 is closer to the desired identity. A careful analysis of the XY4 sequence shows that the deviation from the noisefree propagator during the first two pulses is partially compensated by the third and the fourth pulse. This insight can be iterated \cite{Gullion1990}. After a single application of the first four pulses, one applies another four pulses where the phases $0$ and $\pi/2$ are interchanged, leading to a kind of YX4 sequence. In the language of the rotation operation of the previous section, the $n_z$ term is almost inverted such that $n_{z}(\mathrm{XY4}) - n_{z}(\mathrm{YX4}) = O(\epsilon^2)$. The resulting eight pulse sequence (now called XY8) then exhibits an angle where the leading order is $\epsilon^3$. The corresponding Bloch vector $\vec{n}$ points close to the equator of the Bloch sphere (similar to the CP sequence in Fig.~\ref{fig:Bloch}). One can continue doubling the number of pulses (then called XY16, XY32 and so on) to further increase the power of the leading order in a series expansion in $\epsilon$. Clearly, this only makes sense as long as the amplitude error stays the dominant noise source.

While ``XYn'' sequences only use pulse phases $0$ and $\pi/2$, other sequences further explore the set of possible pulse phases. The Knill pulse, for example, is a five pulse sequence with three different phases. It works similarly to the XY8 sequence in the sense that the first order correction to the ideal propagator (here it is a $\pi$ pulse) is of order $\epsilon^3$ instead of $\epsilon$.

Another way to counter imperfect $\pi $ pulses are composite pulses. The idea is to substitute a single $\pi$ pulse with multiple pulses without time separation and different phases between the pulses. Composite pulses are called ``self-correcting'' because within the pulse block itself, errors in amplitude and detuning are minimized. Recently, a systematic way to derive composite pulses was introduced in \cite{Genov_Correction_2014}. The ansatz propagator is a sequence of noisy $\pi$ pulses where the phases are free parameters. The phases are only restricted to obey a symmetry with respect to an inverted order. In addition to amplitude errors, the authors also treat detuning errors. They are taken into account by replacing $\mathbbm{1}$ by $\exp(-i \gamma \sigma_z)$ in Eq.~(\ref{eq:9}). Instead of directly optimizing the phases by maximizing, for example, the process fidelity to the ideal operation, one first writes the composite pulse in a power series of $\epsilon$. Then, one tries to find conditions on the phases such that the lowest-order contributions vanish. The more pulses are used, the higher the orders of $\epsilon$ that potentially disappear. Remaining parameters are utilized to minimize the contribution for the first nonvanishing coefficient, either optimized for detuning errors or amplitude errors.

We observe that all mentioned sequences can be derived with the method presented in Ref.~\cite{Genov_Correction_2014} (if the inversion symmetry restriction is dropped). The key point is that the procedure of canceling low orders in $\epsilon$ is independent of the value of $\delta$. However, $\pi$ pulses with free time evolution before and after the pulse like in Eq.~(\ref{eq:11}) are mathematically equivalent to $\pi$ pulses with detuning errors $\gamma = \Delta \tau$. For example, it turns out that the Knill pulse is identical to the U5b sequence in \cite{Genov_Correction_2014}, up to a global shift of the phases. As another example, consider a general four-pulse sequence 
$T_4 = U(\varphi_4)U(\varphi_3)U(\varphi_2)U(\varphi_1)$.
It has a vanishing first-order term in $\epsilon$ if the two conditions $\varphi_2 = \frac{1}{2}(\varphi_3 + \varphi_1 + \pi)$ and $\varphi_4 = \frac{1}{2}(3\varphi_3 - \varphi_1 + \pi)$ are fulfilled. One clearly recognizes the XY4 sequence by choosing $\varphi_1 = \varphi_3 = 0$.

It is natural to combine composite pulses with other sequences like XY4 \cite{Souza2011}. Here we use the so-called U5a sequence from Ref.~\cite{Genov_Correction_2014}, which we only trivially modify to have 
\begin{equation}
\label{eq:15}
\begin{split}
U_{\mathrm{U5a}}&(\varphi) = U(\varphi + 4\pi/3)  U(\varphi + \pi/6) \\ &U(\varphi + 5 \pi/3)  U(\varphi + \pi/6) U(\varphi + 4\pi/3),  
\end{split}
\end{equation}

Note that we explicitly introduce time delays between all pulses (see Eq.~(\ref{eq:11}) and Fig.~\ref{fig:schematics} for an example of two consecutive $U_{\mathrm{U5a}}(0)$ pulses). Due to the robustness of composite pulses with respect to detuning errors, the increased stability against amplitude errors is preserved. We now replace the $U(\varphi)$ in CP, Eq.~(\ref{eq:CP}), and XY4, Eq.~(\ref{eq:13}), by $U_{\mathrm{U5a}}(\varphi)$ with the same phases as before. The new sequences consist of ten and 20 pulses and are denoted by U5a:CP and U5a:XY4 \footnote{Our numerical and analytic results show that U5a:CP gives better results than U5b:CP, which is not discussed in this paper. Interestingly, we find that U5a:XY4 and U5b:XY4 behave identically (after the integration over the inhomogeneous broadening). Notice that, having time delays between the pulses, U5b is equivalent to the Knill pulse. Hence our results found for U5a:XY4 should be found for the KDD sequence Ref.~\cite{Souza2011} as well.}, respectively. One can still find the lowest order for the parametrization in Eq.~(\ref{eq:12}). However, the expressions are lengthy and algebraic in $\cos(a \Delta \tau + b)$ (with $a,b$ some constants) which render some of the following calculations unfeasible. We thus only give the scaling of the leading order in $\epsilon$ in Table \ref{table:parameters2}. It is evident that the combined sequences are significantly more robust, in particular U5a:XY4.

\section{DD sequences applied to SNR}
\label{sec:coher-popul-dd}

The sequences we discussed in the previous section are now applied to coherence, population and SNR introduced in Sec.~\ref{sec:QM_performance}. Using the representation of Eq.~(\ref{eq:12}), quite general formulas are found. For the sake of simplicity, we replace the normalized sum over all spins, $N^{-1} \sum_k$, by an integral over some spectral density $p(\Delta)$ for the detuning, $\int p(\Delta) d\Delta = 1$. The unitary $T_k^{\mathrm{DD}}$ is then simply written as $T^{\mathrm{DD}}$ with an implicit dependency on $\Delta$. If not stated differently, we assume $p(\Delta)$ to be normally distributed with zero mean and a full-width-half-maximum (FWHM) equaling the inhomogeneous spin broadening $\Gamma$.

For the following calculations, it is also helpful to work in the Heisenberg picture for the observables $\sigma_q$, $q \in \left\{ x,y,z \right\}$. Generally, one finds $T^{ \mathrm{DD}\dagger} \sigma_q T^{\mathrm{DD}} = \sum_{p \in \left\{ x,y,z \right\}} g_{q p} \sigma_p$. The transformation matrix $g = \left\{ g_{q p} \right\}_{q p}$ is given in the appendix. We further define $G_{q p} = \int g_{q p} p(\Delta) d\Delta$. 

For all sequences, $\alpha$ is supposed to be small, which is guaranteed if $\epsilon \ll 1$. Note that $\alpha$ depends on $\Delta$. We consider two regimes: few repetitions of the sequence $\alpha m \ll 1$ and many repetitions $\alpha m \gg 1$. Note that it depends on $\alpha$ and hence on the specific sequence what few and many means. For $\alpha m \ll 1$, we expand the trigonometric functions in first orders of $\alpha m$. For $\alpha m \gg 1$, we assume that, in any interval $[\Delta,\Delta + \Theta ]$, $\Theta \ll 1$, $\alpha m \mod 2\pi$ takes values uniformly distributed from $0$ to $2\pi$; while, within the same interval, any considered integrand $f(\Delta) \approx \mathrm{const}$. Hence, we approximate $\int f(\Delta) \sin(\alpha m) d\Delta \approx 0$ and $\int f(\Delta) \sin^2(\alpha m) d\Delta \approx \frac{1}{2}\int f(\Delta) d\Delta$ and similarly for other trigonometric functions.

\subsection{Population noise}
\label{sec:population}

Let us first investigate the noise in terms of population as defined in Eq.~(\ref{eq:5}). We start by integrating over the initial phase of the state, $\beta$. By using the integral expression for $\Delta$ and $\left| s \right\rangle\!\left\langle s\right| = \frac{1}{2}(\mathbbm{1} + \sigma_z)$, we find for one repetition
\begin{equation}
\label{eq:14}
\begin{split}
\rho_{\mathrm{\mathit{ss}}} &= \frac{1}{2} \left(1 -  G_{zz} \right) -\frac{1}{N}\\ 
 & = \int \sin^2(\theta) \sin^2 (\alpha /2)p(\Delta)d \Delta + O(N^{-1}),
\end{split}
\end{equation}
where we omit the $O(N^{-1})$ in the following. Ideally, $\rho_{\mathit{ss}}$ should be zero. We see that, as discussed in Sec.~\ref{sec:DD_sequences}, a small angle $\theta$ like in the XY4 sequence guarantees lower bounds on the population. The formulas for many repetitions read 
\begin{equation}
\label{eq:18}
\rho_{\mathit{ss}}\approx  \frac{m^2}{2}\int \sin^2( \theta) \alpha^2  p(\Delta)d \Delta
\end{equation}
for $\alpha m \ll 1$ and 
\begin{equation}
\label{eq:19}
\rho_{\mathit{ss}} \approx \int \sin^2( \theta)  p(\Delta)d \Delta
\end{equation}
for $\alpha m \gg 1$. We observe that in the first limit $\rho_{\mathit{ss}}$ increases quadratically with the number of repetitions, while it is constant in the second limit. We give the analytic expressions for some sequences in table \ref{table:parameters2}. For these results, we consider the limit $\Gamma \tau \gg 1$ for reasons discussed in the following section. One clearly observes an increased suppression of the population noise with increased complexity of the sequence.

\begin{table}[ht]\setlength{\tabcolsep}{0.5 cm}
\centering
\begin{tabular}{l l l} 
\hline\hline 
 Sequence&$\rho_{\mathit{ss}}$ for $\alpha m \ll 1$& $\rho_{\mathit{ss}}$ for $\alpha m \gg 1$ \\ 
\hline 
CP & $m^2 \epsilon^2$ & 1\\ 
XY4 &$\frac{1}{8}m^2 \epsilon^6$& $\frac{1}{2}\epsilon^2$\\
XY8 &$\frac{1}{4}m^2 \epsilon^6$& 1\\
U5a:CP &$\approx 0.038 m^2 \epsilon^6$& n.a. \\
U5a:XY4 &$\approx 0.67 m^2 \epsilon^{18}$& n.a.\\ 
\hline \hline
\end{tabular}
\caption{\label{table:population}
 Lowest-order expansion of Eqs.~(\ref{eq:18}) and (\ref{eq:19}) for several sequences in the limit $\tau \Gamma \gg 1$. For U5a:CP and U5a:XY4, analytic results are not available (n.a.) for $\alpha m \gg 1$.}
\end{table}

\subsection{Phase Coherence}
\label{sec:coherence}

For the coherence, Eq.~(\ref{eq:3}), we first consider the expectation values 
\begin{equation}
\label{eq:20}
\langle \sigma_x \rangle_{\bar{\rho}} =  \frac{1}{\sqrt{N}}(g_{xx} \cos \beta + g_{xy} \sin
  \beta)  - g_{xz} + O(N^{-1})
\end{equation}
and similarly for $\langle \sigma_y \rangle_{\bar{\rho}}$ where again we omit $O(N^{-1})$ corrections in the following. 
Subsequently, one has to square $\langle \sigma_x \rangle_{\bar{\rho}}$ and $\langle \sigma_y \rangle_{\bar{\rho}}$ and integrate over $\beta$. Then, many cross terms from the squaring disappear. Finally, the coherence reads
\begin{equation}
\label{eq:21}
\begin{split}
  \eta_{\mathrm{coh}} &\approx \frac{1}{2}\left( G_{xx}^2 +
    G_{xy}^{2} + G_{yx}^{2} +G_{yy}^2\right) \\ & +
N(G_{xz}^2 + G_{yz}^2).
\end{split}
\end{equation}
Note that $G_{xx}^2 + G_{xy}^{2} + G_{yx}^{2} + G_{yy}^2$ in Eq.~(\ref{eq:21}) is proportional to the coherence of the initial state $| \psi_{\mathrm{in}} \rangle $, while the second part purely comes from coherence induced by imperfect $\pi$ pulses. We thus consider this second part as additional noise. However, this artificial coherence decays within the time scale of the inhomogeneous broadening $\Gamma$, like the coherence from the initial state does. Mathematically, this can be confirmed for all sequences discussed in this paper and for $\alpha m \ll 1$. There, $g_{xz}$ and $g_{yz}$ can be written as a weighted sum of factors $e^{-i a \Delta \tau}$ with $a \neq 0$. The integral $\int e^{-i a \Delta \tau} p(\Delta) d\Delta$ gives $e^{-(a \Gamma \tau)^2/2}$ if $p(\Delta)$ is Gaussian or $e^{-a  \Gamma \tau}$ if $p(\Delta)$ follows a Lorentz distribution. We thus consider a large enough $\tau$ such that $G_{xz}^2 + G_{yz}^2 \ll N^{-1}$.

In Table \ref{table:coherence}, we list the lowest-order expansions of Eq.~(\ref{eq:21}) in the limit $\Gamma \tau \gg 1$ for the DD sequences. Similarly as for the population, the coherence decays with $1-O(m^2)$ for $\alpha m \ll 1$ and approaches a finite value in the limit of many repetitions. Again, we find that more complex sequences preserve coherence better.

\begin{table}[ht]\setlength{\tabcolsep}{0.2 cm}
\centering
\begin{tabular}{l l l} 
\hline\hline 
 Sequence&$1-\eta_{\mathrm{coh}}$ for $\alpha m \ll 1$& $1-\eta_{\mathrm{coh}}$ for $\alpha m \gg 1$ \\ 
\hline 
CP & $m^2 \epsilon^2$ & $\frac{1}{2}$\\ 
XY4 &$\frac{1}{2}m^2 \epsilon^4$& 0\\
XY8 &$\frac{1}{4}m^2 \epsilon^6$& $\frac{1}{2}$\\
U5a:CP &$\approx 0.038 m^2 \epsilon^6$& n.a. \\
U5a:XY4 &$\approx 0.81 m^2 \epsilon^{12}$& n.a.\\ 
\hline \hline
\end{tabular}
\caption{\label{table:coherence}
 Lowest-order expansion of $1-\eta_{\mathrm{coh}}$ from Eq.~(\ref{eq:21}) for several sequences in the limit $\tau \Gamma \gg 1$. For U5a:CP and U5a:XY4, analytic results are not available (n.a.) for $\alpha m \gg 1$.}
\end{table}

\subsection{SNR}
\label{sec:application-snr}

For the analysis of the SNR, one simply has to combine the results for population and coherence. In this section, we only consider the part of the SNR that is influenced by the DD and define the ratio
\begin{equation}
\label{eq:26}
R = \frac{\eta_{\mathrm{coh}} }{\rho_{\mathrm{\mathit{ss}}}} = \frac{1}{2}\frac{G_{xx}^2 + G_{xy}^{2} + G_{yx}^{2} + G_{yy}^2}{\rho_{\mathit{ss}}},
\end{equation}
where the second equation is valid in the limit $\tau \Gamma \gg 1$. In the case of perfect sequences, one has $T^{\mathrm{DD}} = \mathbbm{1}$, which results in $G_{xx} = G_{yy} = 1$ and $G_{xy} = G_{yx} = \rho_{\mathit{ss}} = 0$; thus $R$ diverges. We now discuss $R$ in the presence of amplitude errors in the two limits $\alpha m \ll 1$ and $\alpha m \gg 1$.

\textit{Regime $\alpha m \ll 1$.--- } Clearly, the SNR is either reduced by a decreased signal or an increase of noise. However, comparing Tables \ref{table:population} and \ref{table:coherence}, one sees that for $\alpha m \ll 1$ small increases of $\rho_{\mathit{ss}}$ alters $R$ significantly, where, at the same time, $\eta_{\mathrm{coh}}$ only slightly deviates from one. Thus, the growth of population dominates the behavior of the SNR. It clearly helps if $\vec{n}$ in Eq.~(\ref{eq:12}) points towards the poles (like for XY4), but it is not necessary. It can be equally compensated by strong suppression of the magnitude of $\alpha$, as in the examples of XY8 and U5a:CP with rotation axes close to the equator (see Tables \ref{table:parameters2} and \ref{table:population}). Nevertheless, the improvement from XY4 to XY8 and U5a:CP is not as tremendous as from U5a:CP to U5a:XY4, which again rotates around an axis very close to the poles. 

\textit{Regime $\alpha m \gg 1$.--- } In this regime, a simple expression for the approximate $R$ can be found. For a simpler treatment, we rotate the reference frame of the spin around the $z$ axis to minimize the $n_y$ component and maximize the $n_x$ component of $T^{\mathrm{DD}}$. This is always possible due to the phase averaging done for the coherence and population. Since $\phi$ may depend on $\Delta$, we cannot always find $n_y = 0$ (e.g., XY8). However, at least for the examples discussed in this paper, one can always find $n_y \ll n_x$ for $\epsilon \ll 1$. Given this and the other assumptions mentioned earlier, one finds 
\begin{equation}
\label{eq:27}
R \approx \frac{\left[  \int \sin^2( \theta)  p(\Delta)d \Delta \right]^2}{\rho_{\mathit{ss}}} \approx \rho_{\mathit{ss}} \in [0,1].
\end{equation}
Hence there is a limit for the performance of the DD sequence, which is reached when $\alpha m \gg 1$. Irrespective of the DD sequence, one has $R \lesssim  1$. However, we emphasize again that it strongly depends on the DD sequence when we enter this regime. The sequence determines $\alpha$ and hence the scale $1/\alpha$ where $m$ is considered to be a large number. In other words, for a given sequence, one is limited to a storage time $t = \tau L m \lesssim \tau L/\alpha$. Theoretically, there is no limit in suppressing $\alpha$ when only amplitude errors are considered. As an example, consider XY8, which does not perform significantly better than XY4 when $\alpha m \ll 1$. However, since $\alpha(XY8) \ll \alpha(XY4)$, one is able to perform much more repetitions with XY8 than with XY4 before $R$ drops to $O(1)$ values.

\begin{figure}[htbp]
\centerline{\includegraphics[width=\columnwidth]{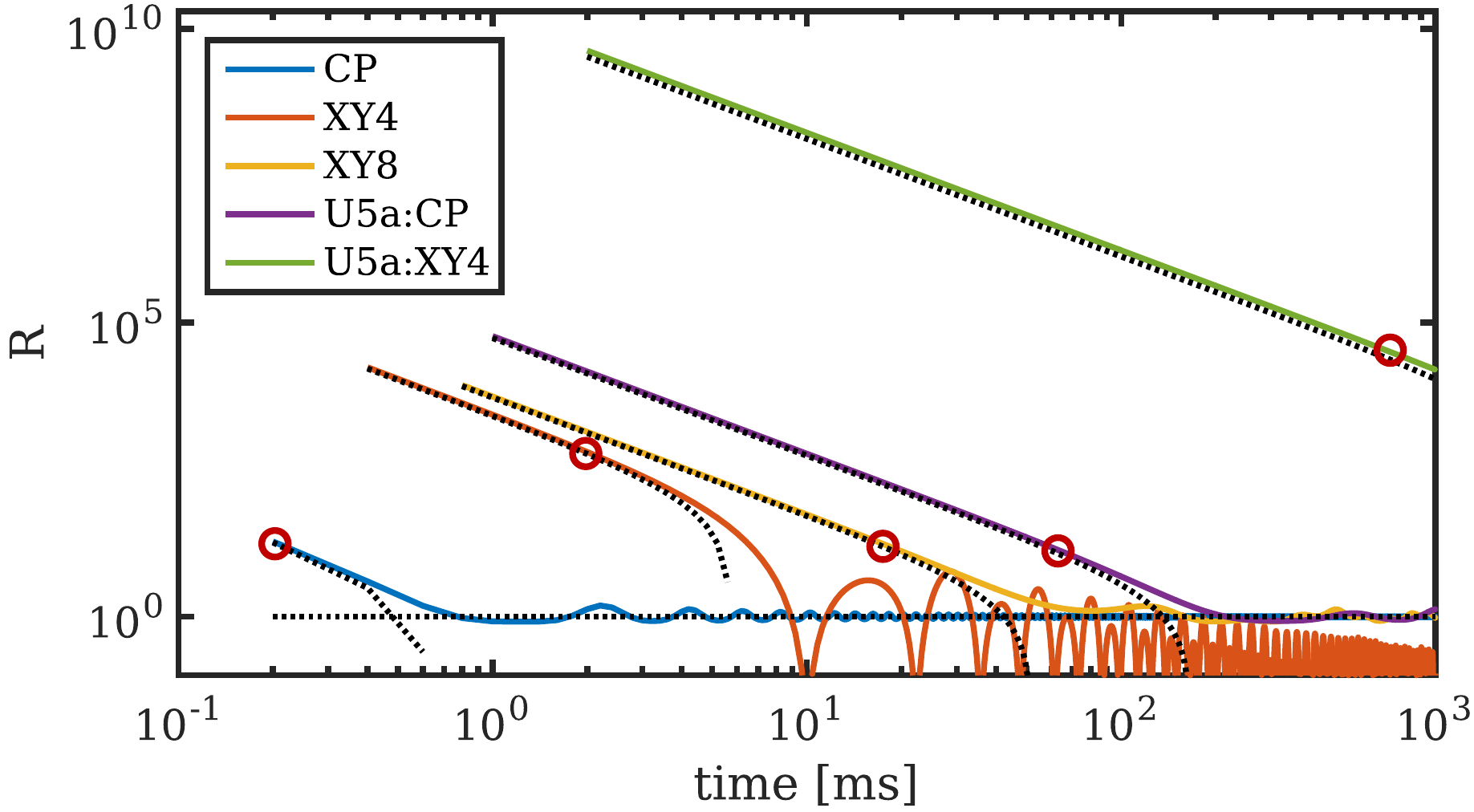}}
\caption[]{\label{fig:SNRInhom} Numerical simulation (solid lines) of $R$, Eq.~(\ref{eq:26}), compared to the analytic results (dotted lines) for different number of repetitions $m$ and for the sequences CP, XY4, XY8, U5a:CP and U5a:XY4 (from bottom to top on the left hand side). The red circles indicate when $\eta_{\mathrm{coh}}$ drops below 0.9. Here, a relatively large amplitude error of $\epsilon = 0.1 \pi$ is chosen to see its influence even for more complex DD sequences. The remaining parameters for the simulation are $\Gamma = 2\pi$ 27 kHz, $\tau = 100 \mathrm{\mu s}$ and $N = 10^{10}$ (taken from the AFC spin-wave experiment performed in a Eu$^{3+}$:Y$_2$SiO$_5$ crystal \cite{Jobez2015}), which guarantees that $N \exp(-\Gamma^2 \tau^2/2) \ll 1$. The analytic expressions, taken from Tables \ref{table:population} and \ref{table:coherence}, fit very well to the results of the simulation.}
\end{figure}

We illustrate these findings and compare the analytic results with a numerical simulation of $R$ in Fig.~\ref{fig:SNRInhom} (see figure caption and appendix \ref{sec:deta-numer-simul} for details on the simulation). Note that the parameter values for $\epsilon$, $\Gamma$, $\tau$ and $N$ are close to the experimental values reported in Ref. \cite{Jobez2015}, where a single XY-4 sequence was used to extend the storage time of an AFC spin-wave memory in Eu$^{3+}$:Y$_2$SiO$_5$. In Fig.~\ref{fig:SNRInhom} one clearly sees that the numerical simulation follows well the analytic approximation for all sequences. The SNR for the simple CP sequence drops most quickly and is the first that reaches the regime $\alpha m \gg 1$. More complex sequences generally perform better. 

\section{The influence of homogeneous broadening and systematic errors in pulse phases}
\label{sec:Numerical}

In this section we additionally take into account homogeneous broadening. This 
is a dephasing process where the detunings are subject to fluctuations in time $\Delta \rightarrow \Delta + \delta(t)$, typically caused by the individual environment of each spin. The effect of time-dependent detunings is that $\pi$ pulses cannot perfectly undo the time evolution, such that $V_{\tau} \Pi_{\varphi} V_{\tau}|_{\epsilon = 0} = \Pi_{\varphi} \exp(-i \nu \sigma_z)$ [cp.~Eq.~(\ref{eq:10})], where $\nu$ is the difference of the time-integrated phase before and after the pulse.

The fluctuations are described by an Ornstein-Uhlenbeck process \cite{uhlenbeck1930theory,Mims1968,Pascual-Winter2012}, which is a stationary, Gaussian and Markovian process. It is characterized by the autocorrelation function

\begin{equation}\label{eq:28}
\langle \delta(t)\delta(t^{\prime})\rangle =\sigma_{\delta}^2 e^{-|t-t^{\prime}|/\tau_{\mathrm{c}}},
\end{equation}
where $\sigma_{\delta}$ determines the mean width of the fluctuations and $\tau_{\mathrm{c}}$ is the correlation time. The time evolution of the Ornstein-Uhlenbeck process is described by a stochastic differential equation that can be exactly simulated \cite{Gillespie_Exact_1996}.

Given amplitude errors and homogeneous broadening, there is a conflict for the optimal choice of $\tau$. To minimize $\nu$, defined above, one normally chooses $\tau$ as small as possible; ideally one takes $\tau \ll \tau_{\mathrm{c}}$. However, smaller $\tau$ implies more pulses for a fixed storage time $t$. Hence, one expects to have more noise through imperfect population inversion. Therefore, the question arises about the optimal value of $\tau$. Based on numerical simulations, we identify three different regimes. First, it is clear that if $\tau \gtrsim \tau_{\mathrm{c}}$, then the DD does not extend the intrinsic dephasing time of the spin ensemble beyond $T_2$ (as shown for the CP sequence without amplitude errors in Ref.~\cite{Pascual-Winter2012}). If $\tau$ is now lowered below $\tau_{\mathrm{c}}$, the phase shift $\nu$ starts to decrease and $T_2^{\mathrm{DD}}$ increases. In the presence of amplitude errors, one might naively think that it is optimal to choose $\tau \lesssim \tau_{\mathrm{c}}$, because further reduction of $\tau$ would induce more population noise without improving the signal.
  However, DD sequences that are more complex than CP profit from smaller $\tau$. As already stated in Ref.~\cite{Genov_Correction_2014}, the basic assumption for the derivation of more complex sequences is that the basic building blocks $U(\varphi)$ [see Eq.~(\ref{eq:11})] are identical (up to the phase $\varphi$). This assumption is violated if within one sequence block (of length $L$) $\delta(t)$ is significantly altered. Hence, in the regime where $\tau \lesssim \tau_{\mathrm{c}} \lesssim \tau L$, the mechanism of error reduction partially fails and the sequence does not perform optimally. Consequently, only in the third regime $\tau \lesssim \tau_{\mathrm{c}}/L$ can one ensure that $\delta(t)$ is approximately  constant over the time scale of one sequence block. In other words, if the correlation time $\tau_{\mathrm{c}} \gtrsim \tau L$, the sophisticated interplay between the pulse phases $\varphi_i$ is fully developed. In the regime $\tau \ll \tau_{\mathrm{c}}/L$ no further stabilization can be expected and more pulses indeed result in more noise. Note that this discussion is still of qualitative character since we did not take $\sigma_{\delta}$ into account. Indeed, our simulations show that $\sigma_{\delta}$ also influences the precise value of the optimal $\tau$.

We present examples of the corresponding simulations in Figs.~\ref{fig:hom_pop} and \ref{fig:SNRhom} (see appendix \ref{sec:deta-numer-simul} for details on the numerical simulation). The total storage time $t = 1 \mathrm{s}$ is fixed and is much larger than $\tau_{\mathrm{c}} = 3.5 \mathrm{ms}$, the correlation time measured for Eu$^{3+}$:Y$_2$SiO$_5$  \cite{Arcangeli2014}. We investigate the impact of various $\tau$ on the population $\rho_{\mathit{ss}}$ in Fig.~\ref{fig:hom_pop} and on the ratio $R = \eta_{\mathrm{coh}}/\rho_{\mathit{ss}}$ in Fig.~\ref{fig:SNRhom}. In particular in Fig.~\ref{fig:hom_pop}, one clearly observes the three regimes. For $\tau \gtrsim 10 ms$, the population is the same for all sequences, because each pulse has no phase relation to the others and hence $\varphi$ does not have any impact. In the regime $\tau \lesssim \tau_{\mathrm{c}}$, the population created by the CP sequence starts to be much larger than for the other sequences, which profit from an increased phase stability over many pulses. When $\tau \lesssim 0.35 \mathrm{ms}$, the population generated by XY8 and U5a:CP again start to increase. With XY4, one has a stabilization of the population (see Secs.~\ref{sec:examples-simple-dd} and \ref{sec:population}), while for U5a:XY4 one could decrease $\tau$ even further.

Note, however, that there are further constraints on $\tau$. As discussed in Sec.~\ref{sec:coherence}, the imperfect population inversion induces unwanted collective emission unless $N \exp(-\tau^2 \Gamma^2/2) \ll 1$ (for a Gaussian profile of inhomogeneous detuning). One should therefore choose $\tau$ large enough to ensure this condition. With the chosen parameters (see also the caption of Fig.~\ref{fig:SNRInhom}), $\tau$ should not go much below 30 $\mu$s. 

\begin{figure}[htbp]
\centerline{\includegraphics[width=1.1\columnwidth]{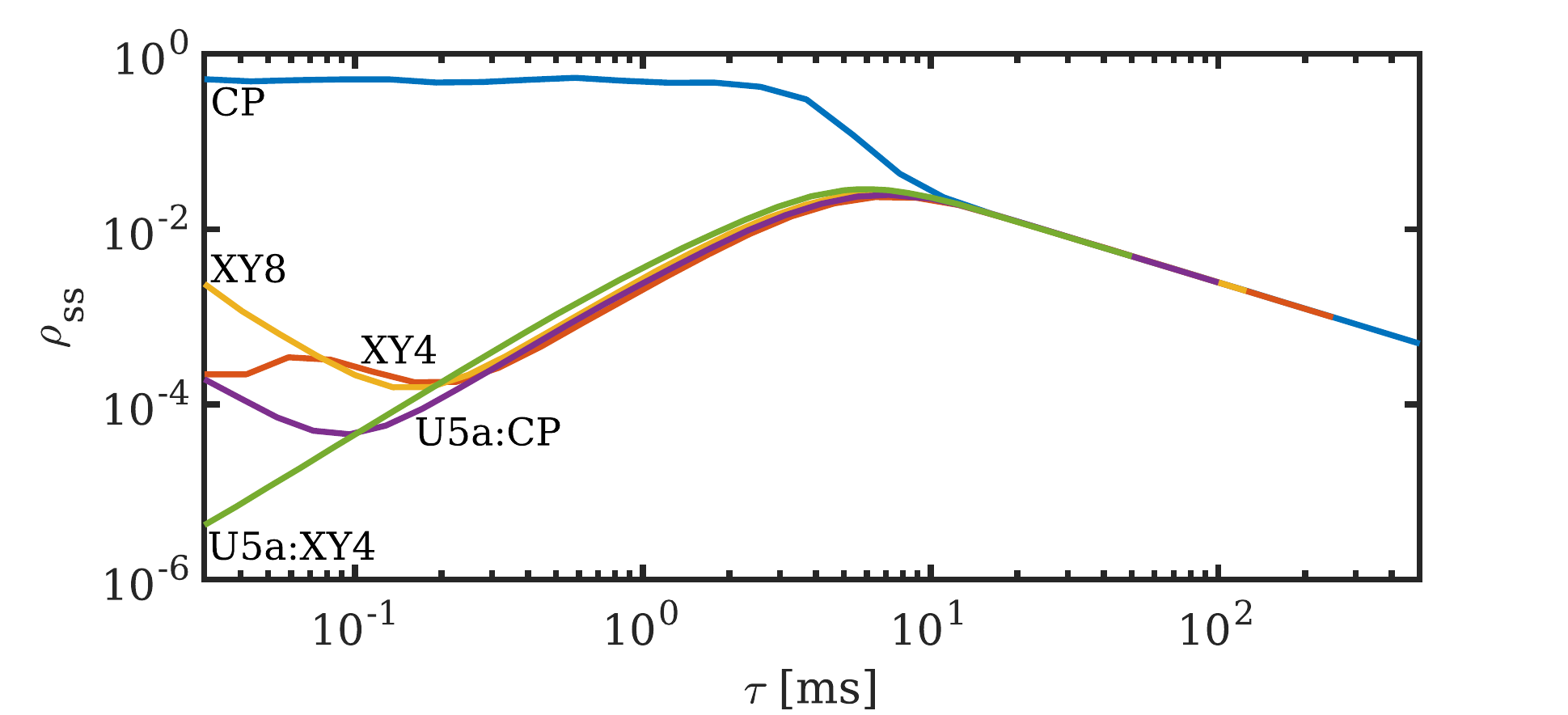}}
\caption[]{\label{fig:hom_pop} Induced noise $\rho_{\mathit{ss}}$ from imperfect $\pi$ pulses in the presence of homogeneous and inhomogeneous broadening for various pulse separations $\tau$. The total storage is fixed $t = 1s$.  The amplitude error is $\epsilon = 0.01\pi$ (ten times smaller than in Fig.~\ref{fig:SNRInhom}). The parameters for the homogeneous broadening are $\tau_{\mathrm{c}} = 3.5 \mathrm{ms}$ and $\sigma_{\delta} = 284 \mathrm{Hz}$ as measured in Eu$^{3+}$:Y$_2$SiO$_5$  Ref.~\cite{Arcangeli2014}. One clearly sees the benefit for more complex sequences over CP in the regime $\tau_{\mathrm{c}}/L \lesssim \tau \lesssim \tau_{\mathrm{c}}$, where $L$ is the length of one sequence (see Table \ref{table:parameters2}).}
\end{figure}

\begin{figure}[htbp]
\centerline{\includegraphics[width=1\columnwidth]{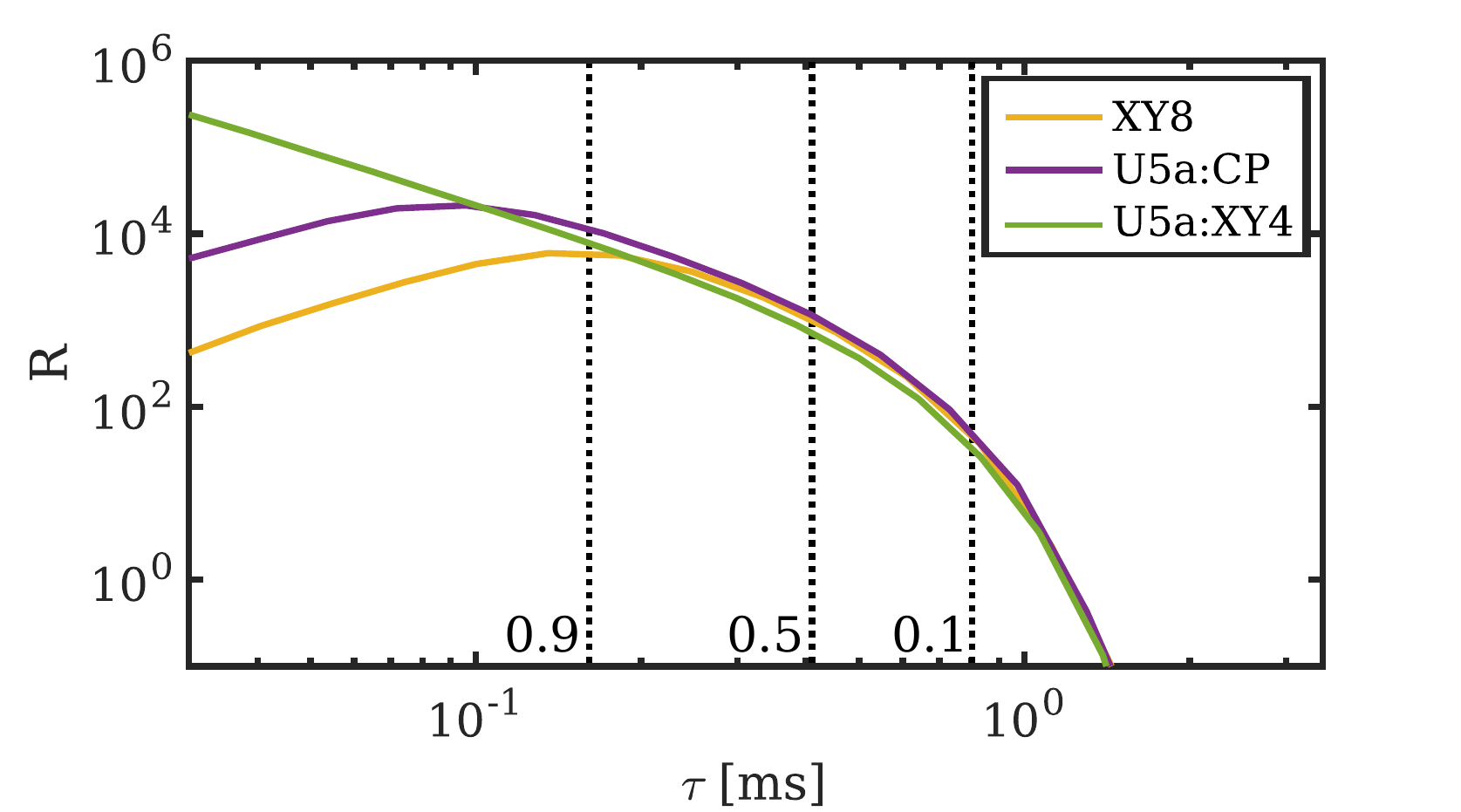}}
\caption[]{\label{fig:SNRhom}  Ratio $R$ for various pulse separations $\tau$. Here, homogeneous broadening, inhomogeneous broadening and amplitude errors are taken into account. We only plot those sequences where we encounter $\eta_{\mathrm{coh}} \geq 0.5$ for at least one choice of $\tau$: XY8, U5a:CP and U5a:XY4 (from bottom to top on the left hand side) are repeated $m$ times such that $m L \tau = 1 \mathrm{s}$. The results are from the same run of the simulator as in Fig.~\ref{fig:hom_pop}. The dashed vertical lines indicates specific values for $\eta_{\mathrm{coh}}$, which is monotonically decreasing with $\tau$ and which is the same for all three sequences.}
\end{figure}

\section{Application in quantum memory experiments}
\label{sec:Conclusions_Outlook}

We now briefly discuss the outlook of applying DD sequences in current quantum memory experiments. As in the rest of the article we focus on the material and experimental parameters used by Jobez \textit{et al.} Ref.~\cite{Jobez2015}. However, a similar analysis is possible for the experiment reported in Ref.~\cite{Rui2015}. In Fig.~\ref{fig:SNRhom} we see that a high ratio $R$ of 10$^3$ or higher could be obtained using either of the three considered sequences, for a pulse separation of around 100 $\mu$s as already used by Jobez \textit{et al.}. Note that the simulations shown in Fig. \ref{fig:SNRhom} includes the homogeneous dephasing process measured in Eu$^{3+}$:Y$_2$SiO$_5$ under similar experimental conditions as in Ref.~\cite{Arcangeli2014}.

To estimate the achievable SNR while storing a single photon, one can use Eq.~(\ref{eq:4}) with $\mu_{in} = 1$. We also consider a memory whose memory efficiency is described by Eq.~(\ref{eq:2.2}), such as an AFC or GEM memory. Let us also assume that additional dephasing is negligible, that is, $\eta_D \approx 1$. In this case we have that SNR = $(1-e^{-\tilde{d}})R$. In the regime $\tilde{d} \gg 1$ it reduces to $R$, hence the values shown in Fig.~\ref{fig:SNRhom} directly give the SNR in the memory output mode. In the regime $\tilde{d} < 1$ the SNR is bounded by $\tilde{d}$. However, since a reasonably efficient memory of, say, $\eta_{M} = \geq 10\%$ would require $\tilde{d} \geq 0.4$, a lower effective optical depth does not strongly affect the SNR either. Obviously the SNR is strongly reduced for very low $\tilde{d}$, but such memories are also very inefficient since $\eta_M \propto \tilde{d}^2$.

In the experiment by Jobez \textit{et al.} Ref.~\cite{Jobez2015} the effective optical depth was $\tilde{d} \approx 1$, which in principle would allow a high SNR after as long as 1 second of storage time using for instance the XY-8 sequence, cf. Fig.~\ref{fig:SNRhom}. In practice the storage time was limited to about 1 ms, using a single XY4 sequence. We believe that the main limitation in that experiment is the multi-level spin states used for storage. Indeed, at close-to-zero magnetic field, due to for instance the Earth B field, the $|g\rangle = |\pm1/2\rangle$ and $|s\rangle = |\pm3/2\rangle$ states in Eu$^{3+}$:Y$_2$SiO$_5$ are split into two closely spaced Zeeman states. The spin echo pulses then simultaneously drive all four spin transitions. The conventional DD sequences studied here only apply to a closed two-level system and do not work well for multi-level systems. 

A potential solution is to apply a large bias B field to be able to spectrally isolate one transition between two spin states. This would, however, also double the number of hyperfine states in both the ground and electronic levels, which possibly could reduce the memory absorption probability (i.e. efficiency) due to difficulties in spin polarizing all ions into a single hyperfine state $|g\rangle$. Cavity-enhancement of the absorption probability can provide one possible solution to this problem \cite{Jobez2014}. But applying a B field would also have the benefit of increasing the spin bath correlation time $\tau_c$, which can reach several seconds for particular field directions as shown by Zhong \textit{et al.} in Ref.~\cite{Zhong2015}. Then the timescale on which a high SNR could be achieved would be 3 orders of magnitude longer than shown in Fig. \ref{fig:SNRhom}, that is 10 minutes or more.

\section{Summary}
\label{sec:summary}

To conclude, we proposed a functional expression for the SNR of spin-ensemble based quantum memories. We subsequently applied the SNR to various known DD sequences and evaluated their performance in the presence of amplitude errors and homogeneous broadening. Our main findings are the following.

Neglecting homogeneous broadening, we parametrized every sequence by a rotation axis $\vec{n}$ and a rotation angle $\alpha$. We identified $\alpha$ as the key parameter of the sequence: The inverse $\alpha^{-1}$ is directly connected to $T_2^{\mathrm{DD}}$, because a reasonable SNR can only be warranted if $m \alpha \lesssim 1$. We confirm that more complex sequences generally have a smaller $\alpha$. Hence, in the regime where the amplitude error dominates, it clearly makes sense to use an elaborate phase relation between the $\pi$ pulses to increase  $T_2^{\mathrm{DD}}$.

In the presence of homogeneous broadening, we found evidence that the optimal pulse delay $\tau$ is in the order of $\tau_{\mathrm{c}}/L$, where $L$ is the length of one sequence block. On first sight, it might be surprising that one profits from reducing $\tau$ from $O(\tau_{\mathrm{c}})$ to  $O(\tau_{\mathrm{c}}/L)$, since more pulses apparently induce more population noise. However, if one has a correlation time that is in the order of the entire sequence block, one can profit from elaborate phase relations between the $\pi$ pulses, which in turn increases the stability of the storage process against amplitude errors. Only if $\tau$ is reduced much below $\tau_{\mathrm{c}}/L$, amplitude errors become dominant and $T_2^{\mathrm{DD}}$ decreases. Note that the precise value for the optimal $\tau$ also depends on $\sigma_{\delta}$.

We also briefly discussed the prospect of applying DD to current quantum memory experiments. Based on our calculations, and the experiments by Jobez \textit{et al.}~\cite{Jobez2015} and Zhong \textit{et al.}~\cite{Zhong2015}, we believe it to be realistic to reach storage times of seconds or even longer in an ensemble-based quantum memory, while achieving a high signal-to-noise ratio in the memory output mode.

\section*{Acknowledgments}
\label{sec:acknowledgments}

We acknowledge financial support from the Swiss National Centres of Competence in Research (NCCR) project Quantum Science Technology (QSIT) and from the CIPRIS project (People Programme (Marie Curie Actions) of the European Union Seventh Framework Programme FP7/2007-2013/ under REA Grant No. 287252)

\appendix

\section{Signal and noise for generic spin states}
\label{sec:signal-noise-generic}

Here, we argue why we can use first- and second-order correlations for phase-averaged coherent states [see Eqs.~(\ref{eq:3}) and (\ref{eq:5})] to measure the influence of free time evolution and DD on the SNR in Sec.~\ref{sec:QM_performance}. We first discuss the figure of merit for spin states after absorption of a single photon. Then, we show that these can be very well approximated by averaged product states.

After the successful storage of a single photon, the optical excitation is transferred to a spin excitation. We assume that the normalized quantum state at this point reads 
\begin{equation}
\label{eq:1}
\left| \mathrm{W}_s \right\rangle = \sum_{k=1}^N c_k\left| g \right\rangle^{\otimes k-1} \otimes \left| s \right\rangle \otimes \left| g \right\rangle ^{\otimes N-k},
\end{equation}
with $c_k \in \mathbbm{C}$. 
We additionally assume that the subsequent time evolution can be written as a product of unitary operations $\mathcal{T}^{\mathrm{DD}} = T_1^{\mathrm{DD}} \otimes \dots \otimes T_N^{\mathrm{DD}}$, where the index indicates that the unitaries are different for every spin. For the SNR, we evaluate the signal in terms of coherence and the noise is given through the population of the state $\left| \mathrm{W}_s^{\prime} \right\rangle =\mathcal{T}^{\mathrm{DD}} \left| \mathrm{W}_s \right\rangle $.

The signal in the photonic mode is determined by the transfer efficiency and $\left| \mathrm{W}_s^{\prime} \right\rangle$. To judge on $\left| \mathrm{W}_s^{\prime} \right\rangle$, we consider the Hamiltonian for the light-matter interaction, which is given by $H_{\mathrm{int}} \propto S^{\Phi}_{-} a^{\dagger} + \mathrm{h.c.}$ where $S^{\Phi}_{-} = \frac{1}{\sqrt{N}}\sum_{k=1}^N e^{i \phi_k} \sigma_{-}^{(k)}$ and $\Phi = (\phi_1,\dots,\phi_N)$. Then, $\eta_{\mathrm{coh}} = \langle S^{\Phi}_{+}S^{\Phi}_{-} \rangle_{\left| \mathrm{W}_s^{\prime} \right\rangle }$ measures the expected photonic signal after re-emission, assuming unit transfer efficiency. %

The noise is similar to evaluate. The spin excitations induced by imperfections during the DD lead to emission of additional photons. Since the transfer of these excitations to the photonic mode is uncorrelated, the cross terms in $\langle S^{\Phi}_{+}S^{\Phi}_{-} \rangle = \frac{1}{N}\sum_{k,l} e^{-i(\phi_k - \phi_{l})}\langle\sigma_{+}^{(k)} \sigma_{-}^{(l)}\rangle$ average to zero and one is left with $\rho_{\mathit{ss}} = \frac{1}{N} \sum_{k=1}^N \langle \left| s \right\rangle\!\left\langle s\right| ^{(k)} \rangle_{\left| \mathrm{W}_s^{\prime} \right\rangle }$.

For a simpler treatment of the problem, we now replace $| \mathrm{W}_s^{\prime} \rangle $ by a product state. Note that $\rho_{\mathit{ss}}$ and $\eta_{\mathrm{coh}}$ are one- and two-body correlation functions, respectively. Hence, it is sufficient to consider one- and two-body reduced density operators $\rho_k = \mathrm{Tr}_{N\backslash k} \left| \mathrm{W}_s^{\prime} \right\rangle\!\left\langle \mathrm{W}_s^{\prime}\right|$ and $\rho_{kl} = \mathrm{Tr}_{N \backslash k,l} \left| \mathrm{W}_s^{\prime} \right\rangle\!\left\langle \mathrm{W}_s^{\prime}\right|$. 
Assuming that we have a rather homogeneous distribution of the excitation (i.e., $|c_k| = O(N^{-1/2})$ for all $k$), it is straightforward to show that the product state $| \phi_{\mathrm{PS}} \rangle = \bigotimes_k T_k^{\mathrm{DD}} \left| \phi_k \right\rangle $ with $\left| \phi_k \right\rangle = \sqrt{1-|c_k|^2} \left| g \right\rangle + c_k e^{i \beta} \left| s \right\rangle $ has approximately the same one- and two-body reduced density operators, after integrating over $\beta$. More precisely, one finds for all $k,l$ that $\rho_k = \int d\beta/(2\pi)T_k^{\mathrm{DD}}\left| \phi_k \right\rangle\!\left\langle \phi_k\right| T_k^{\mathrm{DD}\dagger}$ and $\rho_{kl} = \int d\beta/(2\pi)T_k^{\mathrm{DD}}T_l^{\mathrm{DD}} \left| \phi_k,\phi_l \right\rangle\!\left\langle \phi_k,\phi_l\right|T_k^{\mathrm{DD}\dagger}T_l^{\mathrm{DD}\dagger}  + C_0$ with a correction $\lVert C_0 \rVert = O(N^{-2})$.

We can go one step further and replace $| \phi_{\mathrm{PS}} \rangle $ by a spin-coherent state. For simplicity, we consider the case $|c_k| = \frac{1}{\sqrt{N}}$ in the following. Since we assume a perfect excitation transfer for the absorption, the phases of $c_k$ necessarily have to match with those of $S_{-}^{\Phi}$, that is, $c_k = \frac{1}{\sqrt{N}} e^{-i \phi_k}$. This implies that we can perform a local basis rotation $R_k$ in the $x$-$y$ plane and find that $\langle S_{+}^{\Phi}S_{-}^{\Phi} \rangle_{\left| \mathrm{W}_s \right\rangle } = \langle S_{+}S_{-} \rangle_{\left| \psi_{\mathrm{in}} \right\rangle ^{\otimes N}} + O(N^{-1})$, where $S_{\pm} = \frac{1}{\sqrt{N}} \sum_k \sigma_{\pm}^{(k)}$ and  
\begin{equation}\label{eq:7a}
  |\psi_{\mathrm{in}}\rangle =\sqrt{1-\frac{1}{N}}|g\rangle+e^{i \beta}\frac{1}{\sqrt{N}}|s\rangle.
\end{equation}
There is an apparent complication by having the local unitary $\mathcal{T}^{\mathrm{DD}}$. The basis change $R_k \left| \psi_{\mathrm{in}} \right\rangle  = \left| \phi_k \right\rangle $ and $R_k \sigma_{\pm} R_k^{\dagger} = e^{\mp i \phi_k} \sigma_{\pm}$ does in general not commute with $T_k^{\mathrm{DD}}$, implying that $\langle S_{+}^{\Phi}S_{-}^{\Phi} \rangle_{\left| \mathrm{W}_s^{\prime} \right\rangle } = \int d\beta/(2\pi)\langle S_{+}S_{-} \rangle_{\left| \psi^{\prime} \right\rangle } + O(N^{-1})$ with $| \psi^{\prime} \rangle = \bigotimes_k R_k^{\dagger} T_k^{\mathrm{DD}} R_k \left| \psi_{\mathrm{in}} \right\rangle $. This means that we consider a modified transformation $\tilde{T}_k^{\mathrm{DD}} = R_k^{\dagger} T_k^{\mathrm{DD}} R_k $, which results in an error when taking $\mathcal{T}^{\mathrm{DD}} \left| \psi_{\mathrm{in}} \right\rangle ^{\otimes N} $ instead of $| \mathrm{W}_s^{\prime} \rangle $. However, since $\left| \left\langle \phi_k \right| \left. \psi_{\mathrm{in}}\right\rangle  \right| = 1-O(N^{-1})$, one can always choose $R_k = \mathbbm{1}+r_k$ with $\lVert r_k \rVert = O(N^{-1/2})$. This means that $T_k^{\mathrm{DD}}$ is sufficiently close to $\tilde{T}_k^{\mathrm{DD}}$ such that $T_k^{\mathrm{DD}}\left| \psi_{\mathrm{in}} \right\rangle \approx \tilde{T}_k^{\mathrm{DD}} \left| \psi_{\mathrm{in}} \right\rangle $. Therefore one finds, in very good approximation, $\eta_{\mathrm{coh}} \approx  \int d\beta/(2\pi)\langle S_{+}S_{-} \rangle_{\mathcal{T}^{\mathrm{DD}}\left| \psi_{\mathrm{in}} \right\rangle ^{\otimes N}}$. Introducing $S_x = \frac{1}{2}(S_{+} + S_{-})$ and $S_y = \frac{1}{2i}(S_{+} - S_{-})$ and using the Holstein-Primakoff approximation $[S_{-},S_{+}] \approx \mathbbm{1}$ (omitting $O(1/N)$ corrections), one easily finds that $\eta_{\mathrm{coh}} = \int d\beta/(2\pi)\langle S_x^2 + S_y^2 - \frac{1}{2} \rangle_{\mathcal{T}^{\mathrm{DD}}\left| \psi_{\mathrm{in}} \right\rangle ^{\otimes N}}$. Simple manipulations then lead to 
\begin{equation}
\label{eq:3a}
\eta_{\mathrm{coh}} \approx \int_0^{2\pi} N \frac{d\beta}{2\pi} \left( \langle \sigma_x \rangle_{\bar{\rho}}^2 + \langle \sigma_y \rangle_{\bar{\rho}}^2 \right)
\end{equation}
with $\bar{\rho} = N^{-1}\sum_k T_k^{\mathrm{DD}}\left| \psi_{\mathrm{in}} \right\rangle \left\langle \psi_{\mathrm{in}} \right| T^{\mathrm{DD}\dagger}_k$.

Notice that the above discussion is simpler for $\rho_{\mathit{ss}}$, since one can always choose the local basis rotation $R_k$ such that it does not have any influence on the population. Similar calculations result in 
\begin{equation}
\label{eq:5a}
\rho_{\mathit{ss}} = \int_0^{2\pi} \frac{d\beta}{2\pi} \left\langle s \right| \bar{\rho} \left| s \right\rangle .
\end{equation}

\section{Passive rotations in Heisenberg picture}
\label{sec:pass-rotat-heis}

In Sec.~\ref{sec:coher-popul-dd}, we use the simple, but lengthy formulas for unitary evolution of Pauli operators in the Heisenberg picture. Here, we provide the full expressions. Given any two-level unitary $T^{\mathrm{DD}} = \exp(- i \alpha \vec{n}\cdot \vec{\sigma})$, the Pauli operators transform as $T^{\mathrm{DD}\dagger} \sigma_q T^{\mathrm{DD}} = \sum_{p \in \left\{ x,y,z \right\}} g_{q p} \sigma_p$ where
\begin{widetext}
  \begin{equation}
    \label{eq:2a}
    g = \{g_{q p}\}_{q p}=
    \begin{pmatrix}
\cos 2\alpha + 2n_x^2 \sin^2\alpha & n_z \sin 2\alpha + 2 n_x n_y \sin^2\alpha &  - n_y \sin 2\alpha + 2 n_x n_z\sin^2\alpha  \\
- n_z \sin 2\alpha + 2 n_y n_x \sin^2 \alpha  & \cos 2\alpha + 2n_y^2 \sin^2\alpha &  n_x \sin 2\alpha + 2 n_y n_z \sin^2\alpha \\
  n_y \sin 2\alpha + 2 n_z n_x\sin^2\alpha  &  -n_x \sin 2\alpha + 2 n_z n_y
\sin^2\alpha  & \cos 2\alpha + 2n_z^2 \sin^2\alpha
    \end{pmatrix}
  \end{equation}
\end{widetext}

\section{Details on the numerical simulation}
\label{sec:deta-numer-simul}

Here, we give some details about the numerical simulation whose results are presented in Figs.~\ref{fig:SNRInhom}, \ref{fig:hom_pop} and \ref{fig:SNRhom}. For all three plots, the same algorithm with different parameters was used. Since the homogeneous process is modeled as a stochastic process (Ornstein-Uhlenbeck), the simulation was realized by sampling $N_{\mathrm{sample}}$ spins from the physical state space. The parameters identically used all plots are: $N_{\mathrm{sample}} = 10^5$, $\tau = 100 \mathrm{\mu s}$, $\Gamma = 2 \pi 27 \mathrm{kHz}$ and a total time of one second (i.e., the DD sequence is repeated until to total time elapsed is $1 \mathrm{s}$).

Let us begin with the simplest case: inhomogeneous broadening and amplitude errors. We start with the detunings $\Delta$ randomly chosen from a Gaussian distribution $p(\Delta)$ with zero mean and a standard deviation equaling the inhomogeneous broadening $\Gamma/\sqrt{8 \log 2}$. Then, one of the DD sequences CP, XY4, XY8, U5a:CP and U5a:XY4 is chosen. For this, the time evolution in the Heisenberg picture is simulated for all three Pauli operators. This has to be done for each detuning individually, since the free time evolution depends on $\Delta$. That is, one computes $N_{\mathrm{sample}}$ times the matrix $g = g(\Delta,\epsilon,\tau,m)$. For each repetition $m$ of the DD sequence, the average matrix $G(\Gamma,\epsilon,\tau,m) = \int g(\Delta,\epsilon,\tau,m) p(\Delta)  d\Delta$ is numerically approximated by averaging the action of $g$ over the finite sample. The entries of $G$ are directly used to calculate $|\rho_{\mathit{gs}}|^2/\rho_{\mathit{ss}}$ in Eq.~(\ref{eq:27}).
To see U5a:XY4 (the most stable sequence studied in the paper) being significantly influenced by an amplitude error $\epsilon$, we had to choose a relatively large error of $\epsilon = 0.1 \pi$.

To include homogeneous broadening, the fluctuations of each free evolution step are simulated as described in Ref.~\cite{Gillespie_Exact_1996}. This includes drawing $2 N_{\mathrm{sample}}$ random numbers per time step from a normal distribution. For the amplitude error, we took $\epsilon = 0.01 \pi$. The used parameters of the Ornstein-Uhlenbeck process are $\sigma_{\delta} = 168 \mathrm{Hz}$ for the homogeneous broadening and $t_c = 3.7 \mathrm{ms}$ for the correlation time, which were measured in Ref.~\cite{Arcangeli2014}.

%

\bibliographystyle{apsrev4-1}
\end{document}